\newcommand{\Figdir}{.}

\documentstyle[amssymb,aps,prl,fancyheadings,epsfig,epsf]{revtex}

\begin{document}

\twocolumn[\hsize\textwidth\columnwidth\hsize\csname
@twocolumnfalse\endcsname
\draft
\author{C.D. Batista}
\address{Centro At\'{o}mico Bariloche and Instituto Balseiro, (8400) S.  
C. de Bariloche, Argentina.}
\author{ J. Eroles}
\address{Centro At\'{o}mico Bariloche and Instituto Balseiro, (8400) S.  
C. de Bariloche, Argentina, \\
Los Alamos National Laboratory, Los Alamos NM 87545, USA.}
\author{ M. Avignon}
\address{Centro At\'{o}mico Bariloche and Instituto Balseiro, (8400) S.  
C. de Bariloche, Argentina, \\
 Laboratoire d'Etudes des
Propri\'{e}t\'{e}s Electroniques des Solides (LEPES)\protect\cite{perm}-CNRS -
BP 166, 38042   
Grenoble Cedex 9, France. } 
\author{ B. Alascio}
\address{Centro At\'{o}mico Bariloche and Instituto Balseiro, (8400) S.  
C. de Bariloche, Argentina.}
\date{\today}
\title{Ferromagnetic Polarons in Manganites}
\maketitle
\begin{abstract}
Using the Lanczos method in linear chains we study the
double exchange model in the low concentration limit, including an
antiferromagnetic super-exchange $K$. In the strong coupling limit we find  
that the ground state contains ferromagnetic polarons whose length  
is very sensitive to the value of $K/t$.  We investigate the dispersion
relation,  
the trapping by impurities, and the interaction between these polarons. 
As the overlap between polarons increases, by decreasing $K/t$, the effective
interaction between them changes from antiferromagnetic to ferromagnetic.
The scaling to the thermodynamic limit suggests an
attractive interaction in the strong coupling regime ($J_h > t$)
and no  
binding in the weak limit ($J_h \simeq t$).
\end{abstract}
\vskip2pc]

\narrowtext

\section{Introduction}

The discovery of ``colossal'' magnetoresistance (CMR) \cite{helm} together
with its many unusual properties has received considerable attention lately.
The materials that exhibit this phenomenon are under much experimental
investigation due to their technological applications. These perovskites are
ferromagnetic oxides of the form A$_{1-x}$B$_{x}$MnO$_{3}$ (where
A=La,Pr,Nd; B=Sr,Ca,Ba,Pb) \cite{Jin,Tok}. Experiments have revealed very
rich phase diagrams interpreted in terms of ferromagnetic,
antiferromagnetic, canted, polaronic, and inhomogeneous phases. Charge
ordered phases have also been found\cite{Ramirez1,Coey,Ram2,Kawano,Liu,Bao}
in these compounds.

The phase diagram, as a function of concentration $x$, temperature, magnetic
field, or magnitude of the relation between super-exchange and hopping
interactions is not yet completely understood  for the different compounds. A
clear asymmetry is observed between hole ($x<0.5$) and electron doped ($%
x>0.5 $) regimes. A metallic ferromagnetic phase can be reached by hole
doping of the parent compound LaMnO$_{3}$, by substituting La for divalent
alkalis, Pb, or by stoichiometry changes. On the other side, electron doped
systems exhibit charge ordering or quasi-ordering and antiferromagnetism.

In the hole doped perovskites, the strong correlation between itinerant
carriers and localized spins gives rise to competing magnetic interactions.
This competing interactions plus the effects of disorder and lattice degrees
of freedom, could give rise to inhomogeneities of varying scale: microscopic
polarons, mesoscopic droplets or macroscopic phase separation. A number of
experimental data indicating the different scales in the hole doped regime
can be found in Refs.\cite{Yam,de teresa} for polarons, \cite{Hennion} for
droplets and \cite{Uhara} for phase separation.

Much less is known about the electron doped compounds. This lack of initial
interest can be explained by the fact that manganites such as Bi$_{0.25}$Ca$%
_{0.75}$MnO$_{3}$ or A$_{0.25}$Ca$_{0.75}$MnO$_{3}$ are characterized by
charge ordering phenomena and do not exhibit magnetoresistive effects.
However the recent discovery of new ferromagnetic electron doped manganites,
Ca$_{1-x}$B$_{x}$MnO$_{3}$ and Ca$_{1-x}$Bi$_{x}$MnO$_{3}$ for $x=0.10$ \cite
{Chiba,Troyanchuk}, has increased the interest for those compounds since it
opens the possibility of magnetoresistance effects also in this
concentration limit. Ca$_{0.9}$Bi$_{0.1}$MnO$_{3}$ and Ca$_{0.9}$Eu$_{0.1}$%
MnO$_{3}$ exhibit indeed negative magnetoresistance but with a much smaller
magnitude than hole doped manganites. Long range charge ordering has
recently been observed with electron diffraction in La$_{1-x}$Ca$_{x}$MnO$%
_{3}$ ($x>0.5$) \cite{Ram2,Kawano,Liu,Bao,Murakami,Wollan}. The magnetic
susceptibility has a pronounced inflection at the charge ordering
temperature, resembling that associated with a conventional long range
antiferromagnetic transition\cite{Schiffer}. Neutron scattering measurements
in Bi$_{1-x}$Ca$_{x}$MnO$_{3}$ 
($0.74<x<0.82 $) \cite{Bao} indicate that charge ordering is accompanied by a
structural transition and antiferromagnetic long range order indeed develops
at lower temperature. The nature of spin fluctuations changes from
ferromagnetic to antiferromagnetic at the charge ordering transition \cite
{Bao}. Optical reflectivity studies of the same compound for $T_{co}>T>T_{N}$
have revealed the coexistence of a polaron response and a charge-gap-like
structure in the optical response \cite{Liu}. This two phases behavior is
characterized by domains of ferromagnetic and antiferromagnetic spin
correlations. Magnetization and resistivity measurements in Ca$_{1-x}$Sm$%
_{x} $MnO$_{3}$ for $0<x<0.12$ are interpreted in terms of the existence of
a cluster glass metallic state below some critical temperature and
demonstrate the lack of true ferromagnetism in these electron doped
manganites \cite{Maignan}.

From the theoretical point of view, the pioneering work of de Gennes \cite
{degen} proposed a canted phase to resolve the competition between the
ferromagnetic double-exchange (DE) interaction introduced by the presence of
itinerant holes and the super-exchange (SE) interaction. Recently, several
contributions to this problem have been reported. Arovas and Guinea \cite
{Arovas}, studied this problem using a Schwinger boson formalism to obtain a
phase diagram showing several homogeneous phases and pointing out that phase
separation replaces the canted phase in a large region. Indeed phase
separation appears in several numerical treatments of the problem \cite
{Dagotto1}. In other analytical treatments, more adequate to treat local
instabilities, non-saturated local magnetization states have appeared at
zero temperature \cite{caty}. M. Yu. Kagan {\it et. al.}\cite{Khomsky} have
studied the stability of the canted phases against the formation of large
ferromagnetic 'droplets' containing several particles and they conclude that
the formation of droplets is favored in the ground state. The variety of
results obtained from the different approaches points to the need of
clarifying the picture and testing the results.

The diversity of results in the phase diagram is connected to the two
dominant magnetic interactions acting in these systems: double exchange
arising from the e$_{g}$ orbitals and antiferromagnetic superexchange due
mainly to half-filled t$_{g}$ orbitals. These competing interactions could
give rise to textures of different scales according to the relation between
them.

Here we study a model for {\em diluted compounds} that characterizes them in
terms of a single parameter: the relation between superexchange and hopping
energies. In fact the ionic radii of the intervening dopants varies
substantially the hopping matrix element through the variation of the
Mn-0-Mn angle as discussed in Ref.\cite{Ibarra}.

In this work, we find the low energy quasiparticles and characterize their
structure and dispersion relation in the low concentration limit. These
quasiparticles correspond to the electron followed by a ferromagnetic local
distortion (ferromagnetic polaron) in the antiferromagnetic (AF) background.
The dispersion relation is dominated by $k\rightarrow k+\pi $ scattering due
to the presence of AF order. In order to make a connection with transport
properties, we study the tendency to localization of these polarons in the
presence of impurities and magnetic field.

We also study the interaction between the quasiparticles. For two particles,
the profile of each quasiparticle is practically the same as the one
obtained for only one particle. The effective magnetic interaction between
these quasiparticles is antiferromagnetic for large values of $K$ and
becomes ferromagnetic when the size of the polarons increase over some
critical value. This change in the spin-spin correlation is followed by a
change in the charge-charge interaction. If the size of the ferromagnetic
distortion induced by one particle is larger than some critical value the
other particle could share the same distortion giving rise to a bipolaronic
bounded state.

We include diagonal disorder in the two particles problem in order to
determine how this affects the effective interactions between the polarons.
We find that randomness in diagonal energies induce a distribution of
effective interactions between localized polarons which can be ferro or
antiferromagnetic.

\section{The model}

To render evident the nature of the ground state, we resort to the Lanczos
method, which is free from approximations. The Hamiltonian is simplified to
a single orbital per site, no lattice effects are considered, and we have to
reduce to one dimensional chains. However our results provide a simple
picture that, we presume, can put to test the dilute limit of electron doped
systems. In these systems, the limitations of the model Hamiltonian may not
be as stringent as in the hole doped systems for the following reasons: a) the
lattice structure is more symmetric so Jahn Teller distortions should play a
less important role (this observation is consistent with the fact that hole
doped manganites La$_{1-x}$Ca$_{x}$MnO$_{3}$ are insulating for low
concentrations of dopant while electron doped ones Ca$_{1-x}$Sm$_{x}$MnO$%
_{3} $ are semi-metallic for similar values of $x$ \cite{Maignan}); b) the large
in-site Coulomb repulsion inhibits double occupation so that it may be
possible to describe the physics by the use of a single effective orbital; c)
the antiferromagnetic structure of two inter-penetrating lattices
can be properly described in one dimension.

In order to describe the manganites we consider two degrees of freedom:
localized spins that represent the $t_{2g}$ electrons at the Mn sites, and
itinerant electrons that hop from $e_{2g}$ Mn orbitals to nearest neighbor 
$e_{2g}$ orbitals. The model Hamiltonian includes Hund coupling $(J_{h})$
between localized and itinerant electrons, an antiferromagnetic
superexchange interaction between localized spins $(K)$, a hopping term of
strength $t$ which we will use as energy unit hereafter and an on site ($U$)
and nearest neighbor ($V$) Coulomb repulsions:

\begin{eqnarray}
H & =&-J_{h}\sum_{i}{\bf S}_{i}\cdot {\bf \sigma }_{i}+K\sum_{<i,j>}{\bf S}%
_{i}\cdot {\bf S}_{j} \\
&&+\sum_{<i,j>,\sigma }t_{ij}\left( c_{i\sigma }^{+}\cdot c_{j\sigma
}+h.c.\right) \,  \nonumber \\
&&+V\sum_{<i,j>}n_{i}n_{j}+U\sum_{i}n_{i\uparrow }n_{i\downarrow }  \nonumber
\end{eqnarray}

\smallskip

\noindent where $n_{i,\sigma }=${\bf \ }$c_{i\sigma }^{+}$ $c_{i\sigma }$ ($%
n_{i}=n_{i\uparrow }+n_{i\downarrow }$), and $c_{i\sigma }^{+},$ $c_{i\sigma
}$ creates and destroys an itinerant electron with spin $\sigma $ at site $i$%
, respectively. {\bf \ }${\bf S}_{i}$ and ${\bf \sigma }_{i}$ are the
localized and itinerant spin 1/2 operators at site $i$, respectively. In
order to reduce the Hilbert space we take S =1/2 for the localized spins
instead of S =3/2. It has been shown numerically that, in the absence of
antiferromagnetic coupling, the results for S =3/2 and 1/2 are qualitatively
similar\cite{Dagotto1}. In the absence of AF coupling the Hamiltonian
reduces to the model called Ferromagnetic Kondo Lattice (FKL) studied with
different numerical methods . The reported phase diagram is very similar for
dimensions D=1, 2, 3 and even in infinite dimension showing a ferromagnetic
phase for large $J_{h}/t$ \cite{Dagotto1,Moreo}. In the low electron
concentration limit ($x\sim 1$) this ferromagnetic phase is obtained for any
value of $J_{h}$. In this limit the AF super-exchange coupling between
localized spins ($K$) is needed to get the observed AF phase. This model has
been studied with Monte Carlo method for classical spins and finite
concentration \cite{Moreo}. Recently some results have also been obtained
for a very small value of $K/t$ ($K/t=0.05$)\cite{Dagotto2}. In this paper
we focus on the dilute limit considering the cases of one and two added
electrons. Preliminary results have been reported in reference \cite{Nos1}.

In the following we will concentrate on the case $U=V=0$, unless stated
otherwise. For the manganites, it is believed that the coupling $J_{h}$ is
large. However there are some observations in LaMnO$_{3}$ and CaMnO$_{3}$
which indicate that $J_{h}$ could be not so large. Quoted values refer to $%
J_{h}=1eV$ from optical conductivity data \cite{Okimoto} and bandwidth $%
W=1eV $ \cite{Coey}. In our one dimensional counterpart model this would
mean $J_{h}=4t$. Band structure calculations \cite{Satpathy} also indicate $%
J_{h}=1eV$. A larger value of $J_{h}$ is inferred from scanning tunneling
spectroscopy \cite{Wei}. 
Recent optical conductivity for CaMnO3 gives $1.7eV$ \cite{Jung}, and $2eV$ 
\cite{Sarma}. However $W$ is also larger in the ferromagnetic. Sarma et. al. 
\cite{Sarma} indicate $W=4eV$ $(t=0.3eV)$, giving $Jh/W=1/2$. In one
dimension this would correspond to $J_{h}=2t$. In all cases $U=8-10eV>>J_{h}$%
. For that reasons we will keep the model general considering both the
strong coupling $(J_{h}\sim 10eV)$ and weak coupling $(J_{h}\sim 1eV)\;$%
regimes.

\section{One particle}

In this section we investigate the homogeneity of the solutions for
different values of $K/t$. To this end we calculate the ground state with
one itinerant $e_{2g}$ electron added, for chains of different sizes up to $%
N=20$. For large values of $J_{h}$, the particle modifies substantially the
spin structure in its surroundings forming magnetic polarons. The
distortions of the magnetic structure around the particle can be determined
from correlation functions which mix charge and spin variables. In this way
it is possible to distinguish between polaronic and non-polaronic regimes.

The quasiparticle character of this polaronic distortion becomes evident
from the dispersion relation. The effective mass of the quasiparticle is
inversely proportional to the bandwidth. In this way it is possible to study
the variation of the effective mass as a function of the parameters.

By lowering the diagonal energy in one site it is possible to localize the
polaron. We study the dependence of the localization length and the spin
distortion on the effective mass and on an external magnetic field.

The correlations functions of this section have been calculated using
periodic boundary conditions (PBC).

\subsection{Polaron profile}

\begin{figure}[tbp]
\epsfxsize=3.6in
\epsfysize=4in
\centerline{{\epsfbox{\Figdir/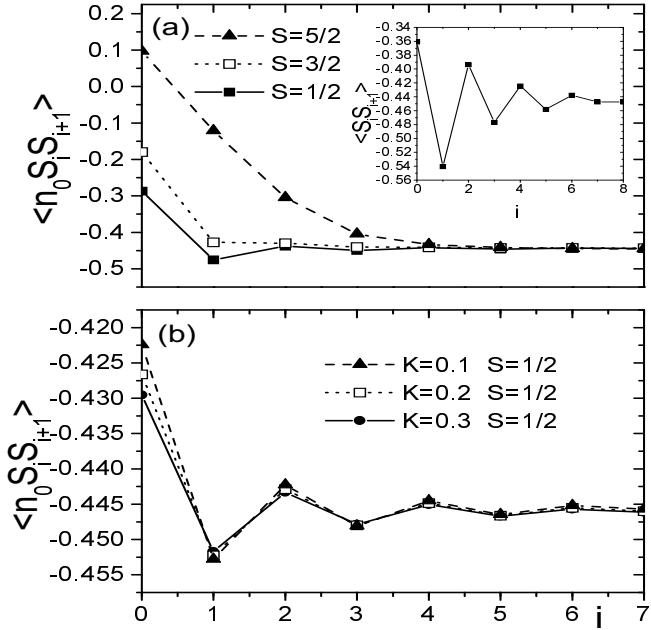}}}
\caption{$a)$ We show the correlation function $<n_{0}S_{j}S_{j+1}>$ for a
16 sites chain, with $J_{h}=10$ and different values of $K$ ($K=0.1$
triangles, $K=0.2$ open squares and $K=0.3$ full squares){\protect \cite{Nos1}}. The maximum value
of $S$ for each value of $K$ is indicated in the figure. One can observe
that the total $S$ and the extension of the magnetic distortion increase as $%
K$ decreases. The oscillatory behavior found at $K=1$ is a consequence of
the weakening of the antiferromagnetic links around the charge. The inset
shows $<S_{j}S_{j+1}>$ for a 16 sites Heisenberg chain where the link at
site zero is a factor $2$ smaller than the rest. $b)$ The same as $a)$ for
the weak coupling regime ($J_{h}=1$).}
\label{fig1}
\end{figure}

In order to determine the polaron profile we calculate $<n_{i}S_{j}S_{j+1}>$
for the ground state. Because of translation symmetry this correlation
function depends only on $\left| i-j\right| $. The results for different
values of $K/t$ and $J_{h}=10t$ are shown in Fig.\ref{fig1}$(a)$ where we
plot $N<n_{0}S_{j}S_{j+1}>$ vs $j$, where $N$ is the number of sites. As it
can be seen in Fig.\ref{fig1}, for large $j$ this correlation function takes
a value very close to the one obtained from the Bethe ansatz solution of the
Heisenberg chain, $<S_{j}S_{j+1}>\cong -0.443$. The extension and the
magnitude of the spin distortion around the particle increases as $K/t$
decreases. For large values of $K/t$, the background of localized spins
remains in a $S=0$ total spin state although it presents distortion in the
spin-spin correlation function due to the itinerant particle. By decreasing $%
K/t$ the total spin increases to $S=1$ (this occurs for $K/t\approx $ $0.3$
in the case of \ref{fig1}(a) for $J_{h}/t=10$) and higher values indicating
a net polarization of the localized spins. In this case, we can refer to a
polaronic regime.

In Fig.\ref{fig1}(b), we show the same correlation function for $J_{h}=t$.
For this value of $J_{h}$ we are already in the weak coupling limit where
the spin distortion is almost null for any value of $K/t$ (notice the change
in the vertical axis scale). The profile of the distortion in this case is,
in fact, very similar to the one obtained in the strong coupling limit ($%
J_{h}\sim 10t$) for large values of $K/t.$ The polaronic regime appears, in
this case, for much lower values of $K/t$.

The oscillations observed in the curve corresponding to $K=0.3$ are also
observed for larger values of $K$. They are a consequence of the weakening
of the antiferromagnetic links around the charge position which produce a
sort of local spin dimerization. Because of the competence between
double-exchange and super-exchange, the effective antiferromagnetic
interaction between the site where the particle is and the nearest neighbor
one is weakened ($\left\langle n_{0}S_{0}S_{1}\right\rangle >-0.443$). The
next spin correlation ($\left\langle n_{0}S_{1}S_{2}\right\rangle $) may be
even stronger than in the undoped case ($\left\langle
n_{0}S_{1}S_{2}\right\rangle <-0.443$) due to this weakening. This forces
the next link ($\left\langle n_{0}S_{2}S_{3}\right\rangle >-0.443$) to be weaker,
and the same kind of reasoning can be applied to the rest. This effect
explains the oscillations of $\langle n_{0}S_{j}S_{j+1}\rangle$ as a function of $j$
for large values of $K/t$. To prove this point we show in the inset the
nearest neighbors spin-spin correlation functions for a Heisenberg chain of
the same size where the link between site zero and one is a factor of two smaller than
the rest.

\begin{figure}[tbp]
\epsfxsize=3.6in
\centerline{{\epsfbox{\Figdir/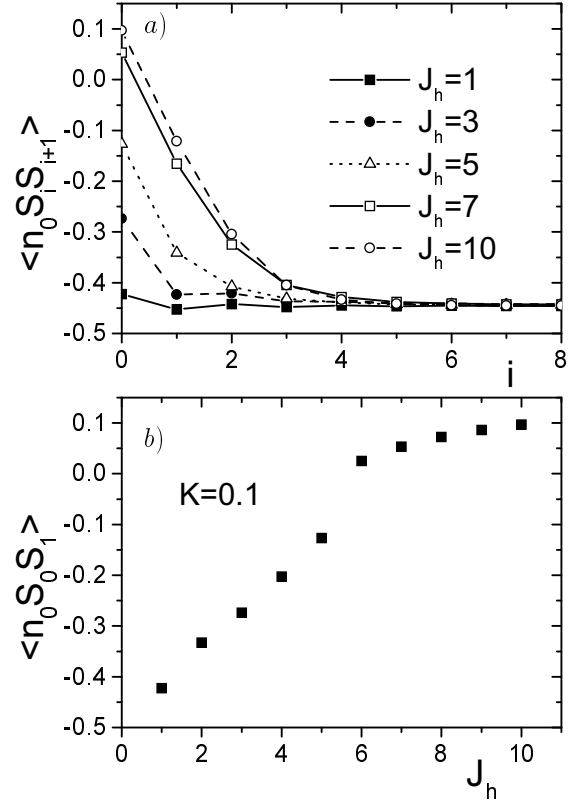}}}
\caption{$a)$ Evolution of the polaron profile from the weak coupling regime
($J_{h}=1$) to the strong coupling limit ($J_{h}=10$). $b)$ Magnetic
correlation between the site where the particle is and the nearest neighbor
as function of $J_{h}$.}
\label{fig2}
\end{figure}

In Fig.\ref{fig2}$(b)$ we show the variation of the polaron profile as a
function of $J_{h}$, for $K/t=0.1$. In this way it is possible to see the
transition from non-polaronic (weak coupling) to polaronic (strong coupling
) regime.

\subsection{Dispersion relation}

As K decreases, it is difficult to find an adequate approximation to
describe the large polaronic distortion. In order to obtain the effective
mass of these polarons, we investigate the dispersion relation for charge
excitations. To this end we calculate the lowest energy state for different
values of the momentum $k =2\pi n/N$ within the subspace where the total
spin is that of the ground state. In Fig.2 from Ref.\cite{Nos1} we show the dispersion
relation scaled to the thermodynamic limit for $K =3$, $J_{h} =100$; $K =1$, 
$J_{h} =10$; and $K =0.3$, $J_{h} =10$.


We start analyzing the dynamics in the regime where $(J_{h}>>K>>t).$ In this
case, the charge moves as a spin one $({\bf \Sigma })$. The effective
hopping resulting from the projection of the hopping term onto the reduced $%
S=1$ Hilbert space is: $tP_{ij}({\bf \Sigma }_{i}{\bf S}_{j}+1/2)$, where $%
P_{ij}$ is the permutation operator between sites $i$ and $j$. We can
picture the movement of the particle, in this limit, as going from a state $%
\downarrow $ $\uparrow $ $\downarrow $ $\Uparrow $ $\downarrow $ $\uparrow $ 
$\downarrow $ to an intermediate state $\downarrow $ $\uparrow $ $\downarrow 
$ $\uparrow $ $0$ $\uparrow $ $\downarrow $, and finally to $\downarrow $ $%
\uparrow $ $\downarrow $ $\uparrow $ $\downarrow $ $\Uparrow $ $\downarrow ,$%
where $\Uparrow $ $(0)$ represents the $S_{z}=+1(0)$ components of the spin $%
S=1$. Thus, in order to move, the charge has to hop to the nearest neighbor,
via a spin flip process, through states that differ in energy by $\Delta
\approx K/2.$ It can be easily verified that the effective hopping of this
process is equal to $t_{ef}=t/\sqrt{2}.$ The dispersion relation given by
this dynamics is: $\Delta /2\pm \sqrt{(\Delta /2)^{2}+4t_{ef}^{2}\cos ^{2}(k)%
\text{ }}$. The expression corresponding to the lower band fits well the numerical 
results for $K=3$
and $J_{h}=100$ as shown in Fig.2 from Ref. \cite{Nos1}. This expression is valid 
in general for a particle moving
in an antiferromagnetic background where scattering between $k$ and $k+\pi $
states dominates the dynamics of the particle.

\begin{figure}[tbp]
\epsfxsize=3.6in
\centerline{{\epsfbox{\Figdir/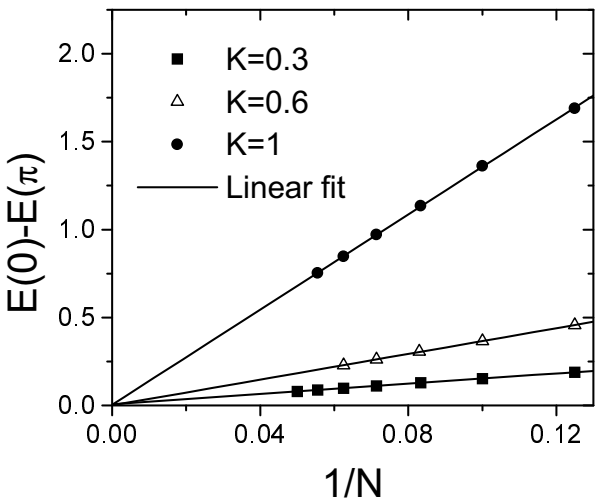}}}
\caption{Scaling of the difference between $k=0$ and $k=\pi$ states for
different values of $K$ and $J_{h}=10$.}
\label{fig4}
\end{figure}

In Fig.\ref{fig4} we show the scaling of the energy difference between $k$
and $k+\pi $ states. This difference must tend to zero as $N$ tends to
infinity due to the reasons explained above. In this way we can check the
quality of the scaling used to calculate the points in Fig.2 from Ref.\cite{Nos1}. The
linear extrapolation goes to zero with an error lower than $5\times 10^{-3}$
for the three values of $K/t$ considered.

In the case $K>>J_{h}>>t$ the spin distortion can be neglected and the
particle propagates in an antiferromagnetic lattice. The Hund interaction
alternates the site energy of the propagating particle so that the
difference between the two sublattices is given by $\Delta =J_{h}(<\sigma
_{j}S_{j+1}>-<\sigma _{j}S_{j}>)\cong J_{h}(<S_{j}S_{j+1}>-<\sigma
_{j}S_{j}>)$ where we approximate $<\sigma _{j}S_{j}>\approx 1/4$ its value
at the triplet state, and $<\sigma _{j}S_{j+1}>\approx $ $<S_{j}S_{j+1}>
=\ln 2-1/4$, the Bethe ansatz value. Using these values we find $\Delta
=0.19J_{h}.$ In this case $t_{ef}$ is equal to $t.$

When $J_{h}>>t\gtrsim K,$ the magnetic distortion around the charge is large
and the effective hopping is dominated by the overlap between the magnetic
distortions about the nearest neighbors sites. This last effect dominates
the polaron effective mass. Therefore, the mass of polarons increases when $%
K $ decreases, as in Fig.2 from \cite{Nos1}, where $t_{ef}$ decreases from
0.75 for $K =1$ to 0.23 for $K =0.3$, showing that the spin distortion
around the charge increases in magnitude and extension when $K$ decreases.

The bandwidth calculated for several values of $K$ (Fig. 3 of Ref.\cite{Nos1}%
) clearly shows two regimes: $K<<t$ and $K>>t$. The first corresponds to a
large magnetic distortion and the second corresponds to a smaller one
according to Fig.\ref{fig1}. The bandwidth goes to zero with $K$ indicating
that the effective mass of the polaron increases continuously by decreasing $%
K$. When $K$ is equal to zero the size of the ferromagnetic distortion
becomes infinite and we can not consider any more this distortion as part of
the quasiparticle structure. The quasiparticle scenario is no longer valid
(it is not possible to associate a magnetic distortion to the electron since
the magnetic distortion is infinite). In this limit, the electron moves in a
ferromagnetic background freely and thus generates a bandwidth equal to the
tight-binding $4t$. In our case, since we have considered finite systems,
this happens for a small, but not zero, value of $K$. This regime was not
shown in Fig. 3 of Ref. \cite{Nos1}.


\subsection{Localization}

\begin{figure}[tbp]
\epsfxsize=3.6in
\centerline{{\epsfbox{\Figdir/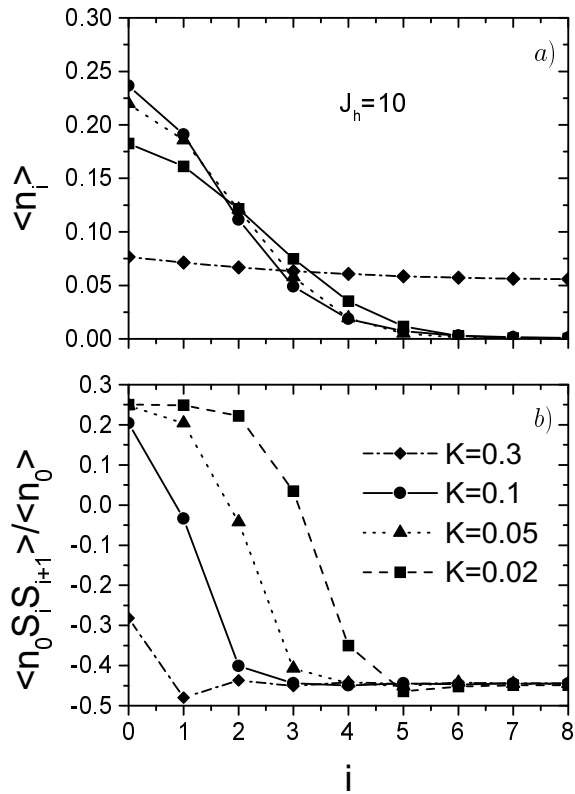}}}
\caption{Charge localization. $a)$ We show $<n_{i}>$ as a function of
position in a 16 sites chain where the energy at site zero is changed from
the rest by $-\epsilon_0$=-0.05 units of $t$. The short localization length
of the lower $K$ curves indicate effective bandwidths of the order of the
energy change. $b)$ We show the extension of the magnetic distortion induced
by the localized particle.}
\label{fig6}
\end{figure}

To test how robust the polaronic description is, we
pin the polaron by changing in $-\epsilon _{0}$ the diagonal energy at site
zero. This may be relevant to the real materials since the doping process
necessarily introduces some disorder. This always localizes the particle in
a linear chain, but the localization length should be very different for
different effective masses. A small $\epsilon _{0}$ localizes much more the
polaron for low values of $K$ than for larger ones. This is shown in Fig.\ref
{fig6}$(a)$ where we plot $<n_{i}>$ around site zero for different values of 
$K$, while fixing $J_{h}=10t$ and $\epsilon_0=0.05$. We also show in Fig.\ref
{fig6}(b) the correlation function $<n_{0}S_{j}S_{j+1}>$ in order to
determine the magnetic distortion associated with each localized state.
While charge localization does not change very much for $K$ between 0.02 and
0.1, the magnetic distortion increases monotonically by decreasing $K$. To
understand this point is useful to take the limit $K/t = 0$ and then slowly
increment $K/t$. For $K/t = 0$, the system is a fully polarized ferromagnet
and thus the ferromagnetic distortion is infinite. In this case, however,
the localization length is finite, although the largest possible. As $K/t $
increases the magnetic distortion will shrink, without affecting the
localization length, at least until they are comparable. In this regime the
magnetic distortion is decreasing while the localization length is not
changing. The difference between the localization and the magnetic
distortion lengths depends on the value of $K$ and $\epsilon _{0}$, and
determines the limit of validity of the quasiparticle concept.


For large values of $K$ the polaronic distortion is very small and the
localization length does not change substantially because the effective mass
of the quasiparticle is similar to the free mass. When $K$ becomes smaller
the localization length starts to decrease as a consequence of the increase
of the effective mass. This behavior remains up to some value of $K$ where
the size of the magnetic distortion is similar to the localization length.
If we further decrease the value of $K$ then the localization length starts
to increase up to the value corresponding to the free case for $K<<$ $%
\epsilon _{0}$, and we loose the quasiparticle picture.

\begin{figure}[tbp]
\epsfxsize=3.6in
\centerline{{\epsfbox{\Figdir/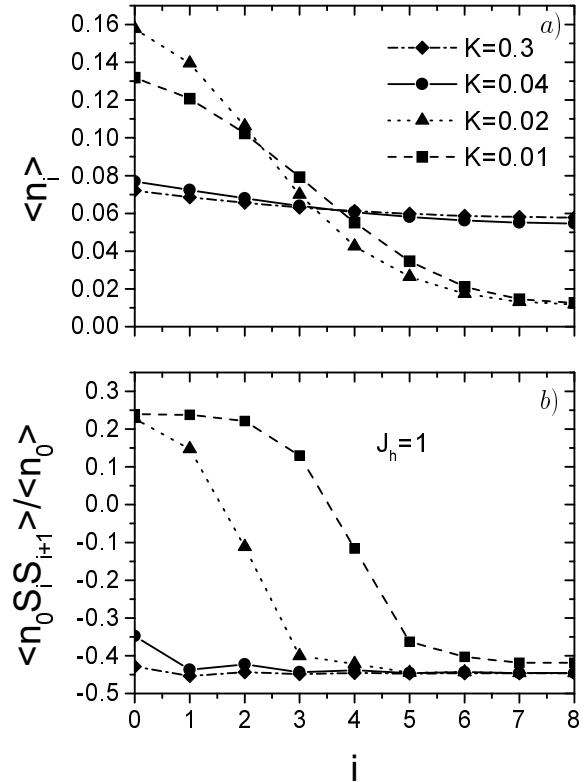}}}
\caption{Same as Fig.{\ref{fig6}} for the weak coupling regime ($J_{h}=1$).}
\label{fig7}
\end{figure}

In Figs.\ref{fig7}$(a)$ and $(b)$ we show the same as in Fig.\ref{fig6} but
for the weak coupling regime ($J_{h} =t$). In this case the localization
length is always larger than in the strong coupling regime as expected. The
size of the magnetic distortion is also smaller. 

\begin{figure}[tbp]
\epsfxsize=3.6in
\centerline{{\epsfbox{\Figdir/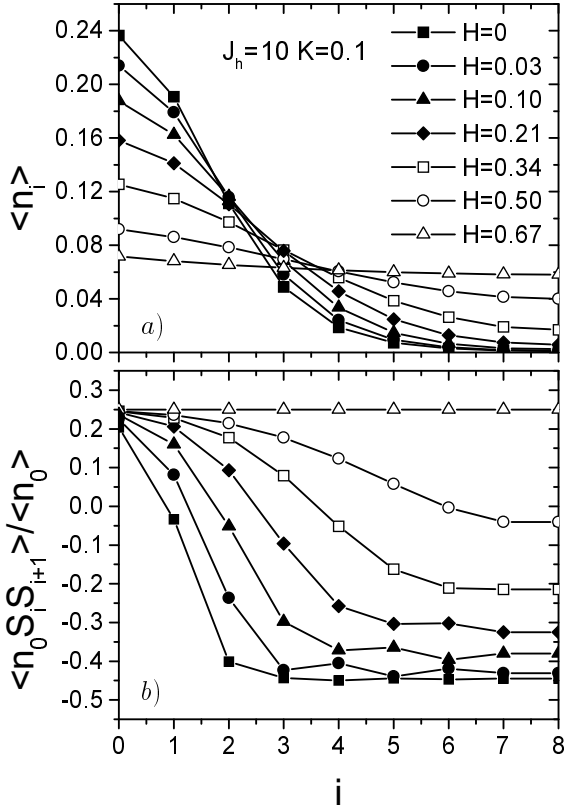}}}
\caption{Effect of magnetic field on charge localization in the strong
coupling limit ($J_{h}=10$). We show $a)$ $<n_{i}>$ for different values of
an external magnetic field for the same chain as in Fig.{\ref{fig4}}; $b)$
the evolution of the magnetic distortion as function of the magnetic field.}
\label{fig8}
\end{figure}

It is interesting to note the differences in the profile of the magnetic
distortion between the localized case (Fig.\ref{fig6}(a)) and the
homogeneous case (Fig.\ref{fig1}(a)). In the localized case the magnetic
distortion is abrupt defining the polaron limits quite sharply (one could
think of a step shape). In the non-localized case the polaron profile decays
more continuously defocusing the polaron limits (one could think of an
exponential-like shape).

In Fig 8$(a)$, we show the change in the values of $<n_{i}>$ for different
magnetic fields in the strong coupling regime ($J_{h}=10t$ and $K=0.1t$). $%
S_{z}$ increases by one between two successive values of $H$ starting from $%
S_{z}=2.5$ for $H=0$. It can be seen that the localization of the polaron
decreases with magnetic field as a consequence of increased effective
hopping between nearest neighbors. A fact that may be important for the
transport properties of these systems since it implies a negative
magnetoresistive behavior for conductivity due to hopping between localized
states\cite{alalmex,Chiba,Troyanchuk}. The size of the magnetic distortion
relative to the background increases with the magnetic field (see Fig.\ref
{fig8}(b)). This could be due to the stronger effect of the magnetic field
on the spins which are close to the boundary of the polaron. Those spins are
forming weak links due to the competition between super and double exchange
so they must have a larger susceptibility.

\begin{figure}[tbp]
\epsfxsize=3.6in
\centerline{{\epsfbox{\Figdir/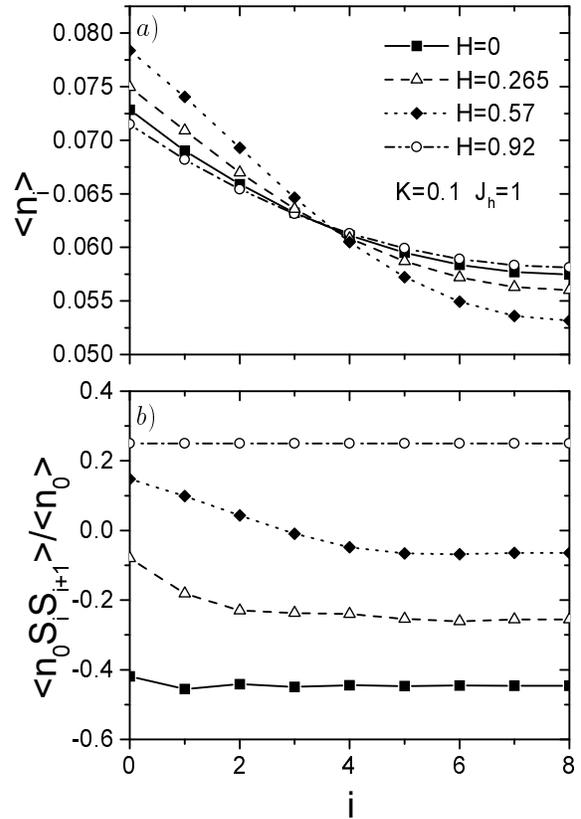}}}
\caption{The same as Fig.{\ref{fig8}} in the weak coupling limit ($J_{h}=1$%
). }
\label{fig9}
\end{figure}

In Fig.\ref{fig9}$(a)$ we show the effect of magnetic field on the
localization for the weak coupling limit ($J_{h}=1,\;K=0.1$). It is clear
from this figure that the magnetoresistive effect is very weak compared with
the strong coupling limit. In addition, the size of the magnetic distortion
is very small (Fig.\ref{fig9}$(b)$) as expected for the weak coupling regime.

\section{Two particles}

In this section we study the interaction between the quasiparticles. To this
end we calculate the magnetic distortion induced by two itinerant electrons,
as well as the binding energy and the charge-charge correlation function. In
this way it is possible to discriminate if the effective interaction is
attractive or repulsive.

\begin{figure}[tbp]
\epsfxsize=3.6in
\centerline{{\epsfbox{\Figdir/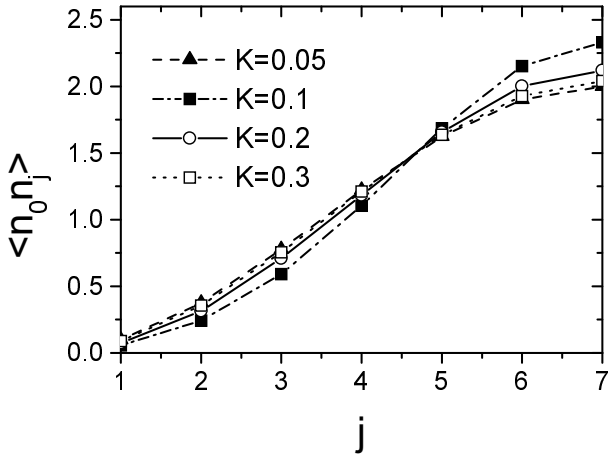}}}
\caption{Charge-charge correlation as a function of distance in the strong
coupling limit ($J_{h}=10$).}
\label{fig10}
\end{figure}

To obtain closed shell conditions in the spinless limit ($K/t=0$
and $J_{h}\rightarrow \infty $) we have taken antiperiodic boundary
conditions (APB). We have only used PBC for the calculation of binding
energies because this quantity also involves the energy of one particle
added, which has been calculated with PBC.

The charge-charge correlation function, $\left\langle
n_{0}n_{j}\right\rangle $ gives information about the character of the
interaction between the polarons. This correlation function is shown in Fig.%
\ref{fig10} for a chain of 14 sites. The maximum always occurs for the
largest possible distance between the particles, what could indicate a
tendency towards repulsion. However in some cases this could just indicate
that the size of the chain is similar or smaller than the mean separation
between the two particles in a bound state. To discriminate between
these two possibilities it is necessary to calculate the binding energy for
different sizes of chains.

\begin{figure}[tbp]
\epsfxsize=3.6in
\centerline{{\epsfbox{\Figdir/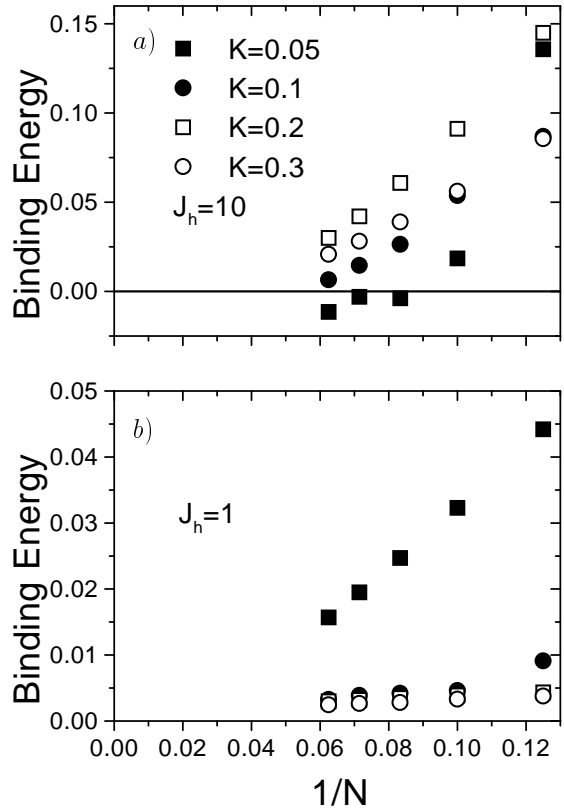}}}
\caption{Binding energy, ($E(2)+E(0)-2E(1)$), as function of the inverse of
the system size in the strong coupling limit ($J_{h}=10$).}
\label{fig11}
\end{figure}

Fig.\ref{fig11}$(a)$ shows the binding energy ($E(2)+E(0)-2E(1)$) as a
function of the inverse of the chain length, in the strong coupling limit.
We find that, although up to the sizes we can compute the interaction is
always repulsive for $K\geq 0.1$, the extrapolation to the thermodynamic
limit seems to give a negative value, suggesting the possible existence of 
a bound state. When $K$ is
low enough ($K=0.05t$) the binding energy is already negative for $N\geq 12$.
The fact that the interaction is repulsive for chains smaller than 12
sites indicates that the mean separation of the bound pair is around six
lattice parameters. In Fig.\ref{fig11}(b) we show the weak coupling limit.
Now for every value of $K$ we get a vanishing binding energy extrapolated to the
thermodynamic limit. This reflects the importance of the double exchange in
the dynamics of the electrons. By increasing the coupling the system seems to cross
between a non-bounding regime into an attractive one.


\begin{figure}[tbp]
\epsfxsize=3.6in
\centerline{{\epsfbox{\Figdir/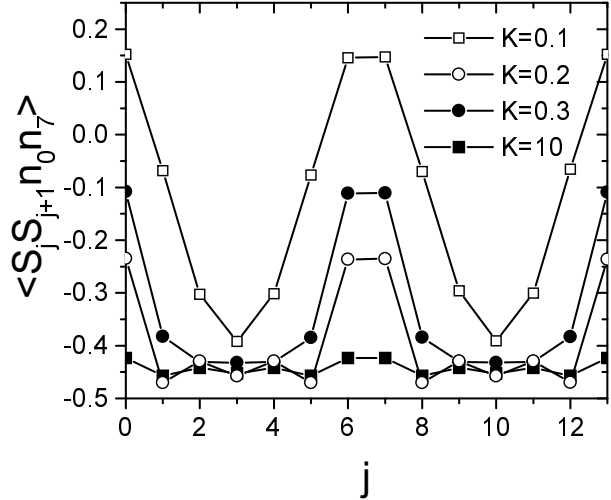}}}
\caption{Profile of the the magnetic distortion with two particles added for
different values of $K$.}
\label{fig12}
\end{figure}

In Fig.\ref{fig12} we show the profile of the magnetic distortion for
different values of $K$ and for the maximum separation between the particles.
As for the one particle case, the size and amplitude of the magnetic
distortion increases as $K$ decrease giving rise to a strong polaronic
regime for $K\sim 0.1t-0.2t$. The two polarons are practically independent
of each other, since their shape is essentially the same as in the single
polaron regime described in the previous section (see Fig.\ref{fig1}). If
the size of each polaron becomes larger than some critical length, one would
expect that the second particle added could gain some magnetic energy by
sharing the distortion created by the first one. This mechanism is similar
to the spin-bag idea proposed by Schrieffer\cite{Schrieffer} to explain the
attractive interaction giving rise to pairing in the high temperature
superconductors. From this consideration, the existence of a bound state
under some critical value of $K$ can be understood.

\begin{figure}[tbp]
\epsfxsize=3.6in
\centerline{{\epsfbox{\Figdir/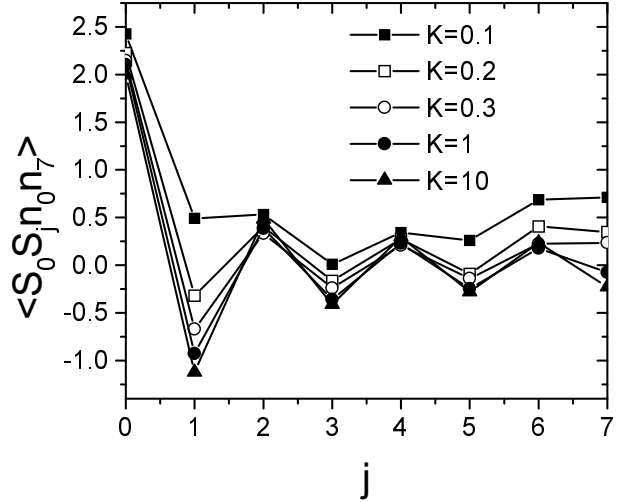}}}
\caption{Spin-spin correlation as function of distance for the largest
separation between the charges.}
\label{fig13}
\end{figure}

\begin{figure}[tbp]
\epsfxsize=3.6in
\centerline{{\epsfbox{\Figdir/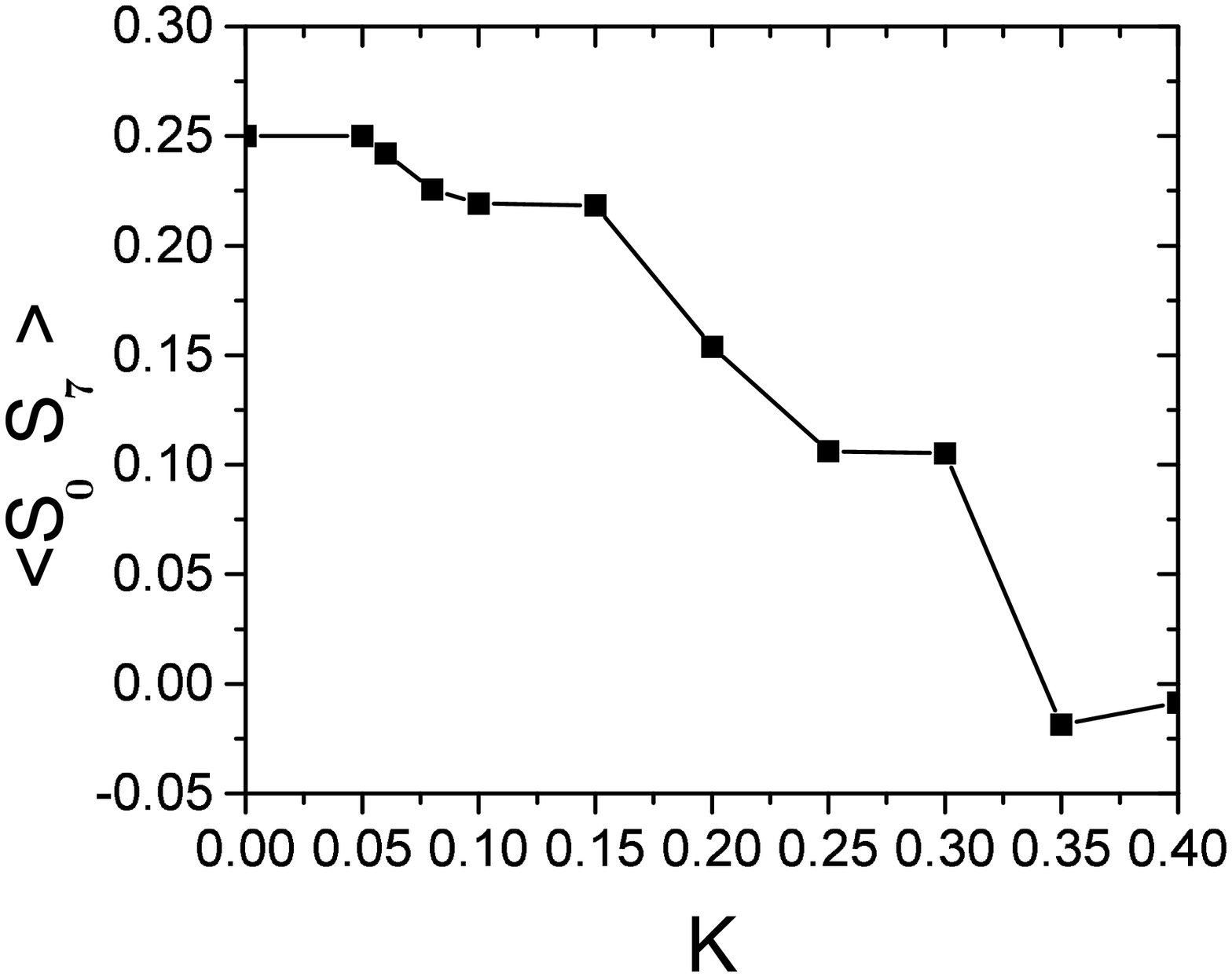}}}
\caption{Spin-spin correlation between the two added electrons for different
values of $K$.}
\label{fig14}
\end{figure}

If the spin-bag picture is correct, the polarons must interact
ferromagnetically when they form a bound state. This can be checked by
calculating the total spin of the ground state and the spin-spin correlation
between the charges. In Fig.\ref{fig13} we show spin-spin correlation as a
function of distance when one particle is at site zero and the other one is
at the longest distance in the chain (site 7 in a 14 sites chain). It is
clear that, as $K$ decreases, the interaction between the two charges
becomes more ferromagnetic. To confirm this result, in Fig.\ref{fig14}, we
show the spin-spin correlation function between the two particles from where
the effective magnetic interaction between the two quasiparticles can be
extracted. It is clear that this interaction evolves from a very small value
for $K=0.4$ to the largest ferromagnetic value for $K=0.1$.

This change in the behavior of the interaction is also obtained in the
charge-charge correlation function (Fig.\ref{fig10}) where, for some
critical value of $K$ between 0.1 and 0.05t, we observe a change in the
variation of the intensity of the maximum with $K$. It increases with
decreasing $K$ from 0.3 to 0.1 and then it decreases practically the same
magnitude between $K=0.1$ and $K=0.05$.

\begin{figure}[tbp]
\epsfxsize=3.3in
\centerline{{\epsfbox{\Figdir/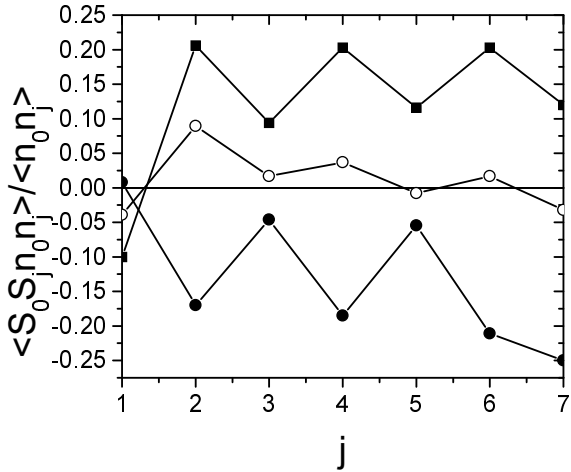}}}	
\caption{Spin-spin correlation function between the two particles as
function of the relative distance. The diagonal energies of sites 0 and 7
have been lowered by $\epsilon _{0}$ in order to pin the two quasiparticles
for $K=0.2$. $\epsilon _{0}=0.01$ (solid squares), $\epsilon _{0}=0.1$ (open
circles), $\epsilon _{0}=1$ (solid circles).}
\label{fig15}
\end{figure}

We also studied the interaction between polarons in presence of disorder. As
it is seen in Fig.\ref{fig6}, lowering of the diagonal energy of one
site not only pins the quasiparticle, but also changes its size. This change
should induce a modification of the effective magnetic interaction between
two pinned quasiparticles. This can be seen in Fig.\ref{fig15} where we show
the spin-spin correlation between the two particles a function of distance
and for different values of $\epsilon_{0}$. The diagonal energy has been
changed by $-\epsilon_{0}$ in both sites 0 and 7. The effective magnetic
interaction evolves from ferromagnetic values for $\epsilon_{0}<0.1$ to
antiferromagnetic for $\epsilon_{0}>0.1$. This can be easily understood from
the reduction in the size of the polaron distortion which is induced by the
increment of the pinning energy $\epsilon_{0}$. As each particle is more
localized, the overlap between polaronic distortions decrease. Each particle
cannot see the magnetic distortion generated by the other and the effective
magnetic interaction is dominated by the superexchange through the localized
spins between the two polarons.

\begin{figure}[tbp]
\epsfxsize=3.6in
\centerline{{\epsfbox{\Figdir/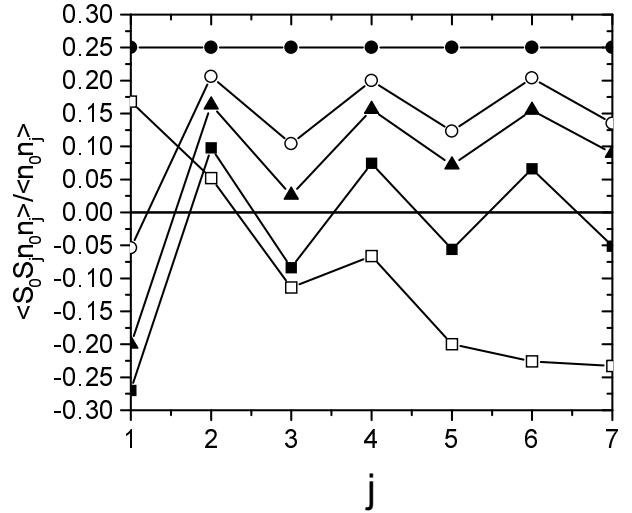}}}
\caption{Same as Fig.{\ref{fig15}}, but for different values of $K$ and $%
\epsilon _{0}=0.05t$. $K=0.05$ (solid circles), $K=0.1$ (open squares), $%
K=0.2$ (open circles), $K=0.3$ (solid triangles), and $K=0.5$ (solid
squares).}
\label{fig16}
\end{figure}

In Fig.\ref{fig16}, we show the variation of the effective magnetic
interaction between two pinned quasiparticles as a function of $K$. In this
case we get a non-monotonic behavior. By decreasing $K$ from 0.5 to 0.2 the
effective interaction becomes more ferromagnetic. But for $K=0.1$ this
behavior changes drastically becoming antiferromagnetic when the particles
are close to each other and ferromagnetic when each one is close to its
pinning center. This indicates the formation of two well defined
ferromagnetic distortions surrounding each center of pining with an
antiferromagnetic effective interaction. For $K=0.05$ this effective
interaction becomes ferromagnetic and we get a uniform ferromagnetic state
(the size of the induced distortion is equal to the size of the system).

Localizing the particle reduces the overlap between polarons, the system
becomes more inhomogeneous and the effective interaction between the
polarons evolves from ferromagnetic to antiferromagnetic. For example, for $%
K=0.1$, $\langle S_{0}S_{7}n_{0}n_{7} \rangle$ is 
positive with $\epsilon _{0}=0$ (Fig.\ref{fig13}) and becomes negative 
for $\epsilon _{0}=0.05$ (Fig.\ref{fig16}).

Finally we have studied how sensitive are the results to the inclusion of $U$
and $V$ in the strong coupling regime. We do not find any significative
change in the physics due to the presence of these interactions. As we are
studying the dilute limit, this interactions should not be so important as
for higher concentrations. Even though, if both particles are at the same
site, they must be in a singlet state due to the Pauli principle. This state
has a very large energy ($\sim J_{h}$) in the strong coupling limit ($%
J_{h}=10$) giving rise to a very small double occupation in the ground
state. This explains the poor role of $U$. To explain the weak effect of $V$
it is useful to take a look at Fig.\ref{fig10} where it can be seen that the
occupation of adjacent sites for the two particles is very small.

\section{Conclusions}

In summary, we have investigated the possibility of strong magnetic
distortions induced by carriers in the ground state of the DE-SE model
Hamiltonian. The model describes chains of localized spins coupled
antiferromagnetically and interacting ferromagnetically with itinerant
electrons. We have studied in detail the case with one and two electrons.
Assuming the model adequately describes the physics of electron doped
manganites, the results presented here point to a picture of these systems
where heavy polarons dominate the magnetic and transport properties. Their
masses depend strongly on the relation between the hopping energy and the
antiferromagnetic superexchange interaction. Clearly, the doping itself will
localize the polarons so that transport will result from hopping between
pinned sites. Negative magnetoresistance should appear as a consequence of
the decrease of the pinning energy with magnetic field\cite{alalmex}.

We have done scaling on the binding energy of two polarons. Two regimes can
be identified depending on the coupling ($J_h$). In the strong coupling
case, although only for small values of $K$ we find a negative binding energy ($K
\leq 0.05$ and $N\geq12$), the scaling to the thermodynamic limit is
negative for every $K$ we have studied, indicating the possible existence of 
a bound state. In the weak coupling case, for most values of $K$, the scaling to the 
thermodynamic limit is zero, and all values are positive. However, as
the charge-charge correlation function (see Fig.\ref{fig10}) is always peaked
at the largest possible distance between the particles, it is necessary to
confirm our scaling results in larger systems. 

The existence of a bound state can be understood from the ferromagnetic
distortion of the polarons. Below some critical value of $K$ this
ferromagnetic distortions of two neighbor polarons overlap and the
charge-charge interaction becomes attractive. This scheme resembles the
spin-bag picture proposed by Schrieffer to explain the pairing mechanism in
the High $T_{c}$ compounds. The effective magnetic interaction between
polarons also changes from antiferromagnetic to ferromagnetic by decreasing
the value of $K$.


We have also studied the effect of localization on the effective interaction
between the quasiparticles. The effect of localization is not only to pin
the quasiparticles but also to change the shape of the ferromagnetic
distortion. As the shape and size of this distortion is clearly connected
with the spin-spin and charge-charge effective interaction we get that the
amount of disorder play a very relevant role in the low energy physics of
these quasiparticles. Depending on the pining energies, ferromagnetic and
antiferromagnetic interactions could be generated. In real systems
frustration induced by such competing interactions could give rise to spin
glass behavior. In this way, the interplay between diagonal disorder and
effective magnetic interactions between polarons could explain the cluster
glass character of the metallic phase found by Maignan et. al. \cite{Maignan}
in Ca$_{1-x}$Sm$_{x}$MnO$_{3}$ for $0<x<0.12$ .

Finally, we would like to point out that the order of oxygen vacancies in %
CaMnO$_{3-\delta }$ makes real the possibility of one dimensional electron
paths in these materials \cite{mate}. We hope that our results will
stimulate more experimental and theoretical investigations on the electron
doped manganites.

\begin{center}
{\it Acknowledgments}
\end{center}

Two of us (C.D.B. and J.E.) are supported by the Consejo Nacional de
Investigaciones Cient\'{i}ficas y T\'{e}cnicas (CONICET). B. A. is partially
supported by CONICET. M.A. gratefully acknowledges support by Universidad
Nacional de Cuyo during his stay at Instituto Balseiro. We would also like
to acknowledge support from the 'Fundacion Antorchas' and the Program for
scientific cooperation between France and Argentina ECOS-SECyT A97E05. This
research is also supported by the Department of Energy under contract
W-7405-ENG-36.

\end{document}